\documentstyle[epsfig,aps,pre,multicol]{revtex}
\tightenlines
\def\beq{\begin{equation}}
\def\eeq{\end{equation}}
\def\beqa{\begin{eqnarray}}
\def\eeqa{\end{eqnarray}}
\def\d{\text{d}}

\begin{document}
\draft
\title{Nonlinear viscosity and velocity distribution function in a simple
longitudinal flow}
\author{Andr\'es Santos\cite{email}}
\address{Departamento de F\'{\i}sica,
Universidad de Extremadura,\\
E-06071 Badajoz, Spain}
\date{\today}
\maketitle
\begin{abstract}
A compressible flow characterized by a velocity field $u_x(x,t)=ax/(1+at)$ 
is analyzed by means of the Boltzmann equation and the 
Bhatnagar-Gross-Krook kinetic model. 
The sign of the control parameter (the longitudinal deformation rate $a$) 
distinguishes between an expansion ($a>0$) and a condensation ($a<0$) 
phenomenon. The temperature is a decreasing function of time in the former 
case, while it is an increasing function in the latter.
The non-Newtonian behavior of the gas is described by a dimensionless 
nonlinear viscosity $\eta^*(a^*)$, that depends on the dimensionless 
longitudinal rate $a^*$. The Chapman-Enskog expansion of $\eta^*$ in powers 
of $a^*$ is seen to be only asymptotic (except in the case of Maxwell 
molecules).
The velocity distribution function is also studied. At any value of $a^*$, 
it exhibits an algebraic high-velocity tail that is responsible for the 
divergence of velocity moments. For sufficiently negative $a^*$, moments of 
degree four and higher may diverge, while for positive $a^*$ the divergence 
occurs in moments of degree equal to or larger than eight.
\end{abstract}
\section{Introduction}
\label{sec1}
One of the most challenging problems in nonequilibrium statistical mechanics
is the understanding of transport properties in fluids beyond the scope of
the Navier-Stokes (NS) constitutive equations.
As part of the NS constitutive equations, Newton's law establishes a {\em
linear\/} relationship between the irreversible momentum flux and the
velocity gradients, namely
\beq
P_{ij}=p\delta_{ij}-\eta_{\text{NS}}\left(\frac{\partial u_i}{\partial x_j}+
\frac{\partial u_j}{\partial x_i}-\frac{2}{d}\nabla\cdot{\bf
u}\delta_{ij}\right)-\zeta_{\text{NS}}\nabla\cdot{\bf
u}\delta_{ij},
\label{1.1}
\eeq
where $P_{ij}$ is the pressure tensor, $p=(1/d)\text{Tr}{\sf P}$ is the
hydrostatic pressure, $d\geq 2$ being the dimensionality of the system, ${\bf 
u}$
is the flow velocity, $\eta_{\text{NS}}$ is the shear viscosity, and
$\zeta_{\text{NS}}$ is the bulk viscosity \cite{GM84}.
The linear law (\ref{1.1}) only holds for small hydrodynamic gradients,
i.e., when the typical distances over which the hydrodynamic quantities
change are much larger than a characteristic microscopic length (such as the
mean free path in the case of gases). Otherwise, Eq.\ (\ref{1.1}) no longer
holds, a situation  usually characterized by the introduction of a
generalized or {\em nonlinear\/} viscosity that depends on the hydrodynamic
gradients \cite{E98}.

The nonlinear viscosity has been extensively studied in the so-called
uniform {\em shear\/} flow, which is characterized by a linear velocity
field $u_x=a y$ and uniform density and temperature
\cite{DBS86,EM90,L97,L98,SMDB98}. This is an example of an incompressible
flow \cite{GM84}, since $\nabla\cdot{\bf u}=0$.
Recently, some attention has been devoted to viscous {\em longitudinal\/}
flows of the form ${\bf u}({\bf r},t)=u_x(x,t)\widehat{\bf x}$
\cite{GK96,KDN97,UP98,KDN98,UG99}.
The simplest example of such compressible flows is characterized by a linear
velocity profile, i.e., $u_x(x,t)=A(t)x$, and uniform density $n$ and
pressure tensor ${\sf P}$ \cite{TM80,G95}. In that case, the balance
equations for mass and momentum read
\beq
\frac{\partial n}{\partial t}=-nA,\quad \frac{\partial (nA)}{\partial
t}=-2nA^2,
\label{1.2}
\eeq
whose solution is
 \beq
A(t)=\frac{a}{1+at}, \quad n(t)=\frac{n_0}{1+at},
\label{1.3}
\eeq
where  $a$ is an arbitrary constant that
represents the (initial) longitudinal deformation rate and $n_0>0$ is the
initial density.
In this case, Newton's law (\ref{1.1}) becomes
\beq
P_{xx}=p-\left(2\frac{d-1}{d}\eta_{\text{NS}}+\zeta_{\text{NS}}\right)A.
\label{1.4}
\eeq
This simple flow is known as {\em homo-energetic
extension\/} and, along with the uniform shear flow, is a particular case of
a more general class of homo-energetic affine flows characterized by
$\partial^2 u_i/\partial x_j\partial x_k=0$ \cite{TM80}.
In the flow defined by Eqs.\ (\ref{1.3}) the longitudinal deformation rate
$a$ is the only {\em control\/} parameter determining the departure of
the fluid from its equilibrium state, thus playing a role similar
to that of the shear rate in the uniform shear
flow state. On the other hand, in contrast to the uniform shear flow, the
sign of $a$ plays a relevant role and defines two distinct situations.
The case $a>0$ corresponds to a
progressively more slowly {\em expansion\/} of the gas from the plane $x=0$ into 
all
of space. Given a layer of width $\delta$, the flux of particles leaving 
the layer exceeds the flux of incoming particles by $na\delta/(1+at)$ and, 
as a consequence, the number of particles inside the layer decreases 
monotonically with time. As time progresses, the system becomes more 
and more rarefied until no particles are left in the long-time limit, i.e.,
$\lim_{t\to\infty}n(t)=0$.
On the other hand, the case $a<0$ corresponds to a progressively 
more rapidly
{\em condensation\/} of the gas towards the plane $x=0$.
The latter takes place over a {\em finite\/} time period $t=|a|^{-1}$. 
However, since the collision frequency rapidly increases with time, 
the finite period $t=|a|^{-1}$
comprises an {\em infinite\/} number of collisions per particle (see below).

 Equations (\ref{1.3}) apply regardless of the initial density of the fluid. On 
the other hand, a kinetic description based on the Boltzmann equation is valid 
only for densities such that $n\sigma^d$ is much smaller than 1, where $\sigma$ 
is a characteristic distance measuring the effective size of the molecules. Let 
us call $n_\sigma\sim \sigma^{-d}$ a characteristic density beyond which 
noticeable deviations from the Boltzmann equation can be expected. Thus, even if 
$n_0\ll n_{\sigma}$, there exists a finite time 
$t_{\sigma}=|a|^{-1}(1-n_0/n_{\sigma})$ beyond which the Boltzmann description 
ceases to be applicable in the case $a<0$. This time $t_{\sigma}$ can be made 
arbitrarily close to the maximum time $|a|^{-1}$ by formally taking the limit 
$n_0/n_{\sigma}\to 0$.

The aim of this paper is to carry out a detailed and self-contained analysis
of the nonequilibrium behavior of a dilute gas under the longitudinal flow
characterized by Eqs.\ (\ref{1.3}), for arbitrary sign and magnitude of the
control parameter $a$. The study is performed by using the tools of
kinetic theory, namely the Boltzmann equation and the Bhatnagar-Gross-Krook
(BGK)  kinetic model, and  deals with the nonlinear viscosity, as well as
with more general velocity moments and the velocity distribution function.
Most of the results are derived for arbitrary dimensionality and for the
general class of repulsive potentials of the form $r^{-\mu}$ with $\mu\geq
2(d-1)$.
The Boltzmann equation for the problem is considered in Sec.\ \ref{sec2}.
Since the density is known, cf.\ Eqs.\ (\ref{1.3}), one can focus on the
probability distribution of velocities. In addition, the distribution
becomes uniform when the velocities are referred to the local Lagrangian
frame moving with the flow velocity $u_x(x,t)$. As a consequence, the 
original problem can be mapped onto that of a
uniform system with a stationary density and subject to the action of a
non-conservative driving force; also, there is a nonlinear relationship
between the time variables in the original and  the equivalent systems.
To proceed further, the Maxwell interaction, $\mu={2(d-1)}$, is assumed and 
the
time evolution of the pressure tensor is exactly obtained. The long-time
behavior allows one to identify the nonlinear viscosity as a function of the
longitudinal deformation rate.
When the velocities are scaled with the (time-dependent) thermal velocity,
the distribution obeys in the  long-time limit a steady-state Boltzmann
equation with the addition of a second non-conservative force playing the
role of a thermostat.
The exact fourth-degree velocity moments are then derived as functions of
the longitudinal rate and are seen to diverge in the case of condensation
for states sufficiently far from equilibrium.
The picture is complemented in Sec.\ \ref{sec3} by the solution of the BGK
kinetic model.
In the case of Maxwell molecules, the distribution function exhibits an
algebraic high-velocity tail that is responsible for the divergence of the
 moments. The solution predicts that moments of sufficiently high degree can
 also diverge in the case of expansion. In addition, the
 distribution function becomes infinite in the limit of zero
 velocity in far from equilibrium states.
Section \ref{sec3} also analyzes the nonlinear viscosity for more general
repulsive potentials, including hard spheres. Evidence is given about the
non-convergent (but asymptotic) character of the Chapman-Enskog expansion,
 Maxwell molecules being an exception. On the other hand, the nonlinear 
 dependence of
the generalized viscosity is practically insensitive to the interaction
potential in the case of positive deformation rates. For negative rates,
however, the influence of the potential is not small, especially near the
maximum value of the viscosity.
The paper ends with a summary of the main conclusions in Sec.\ \ref{sec4}.

\section{Boltzmann description for Maxwell molecules}
\label{sec2}
Let us consider a dilute gas subject to the homo-energetic extension flow
described in the previous Section. All the relevant information is contained
in the one-particle velocity distribution function $f(x,{\bf v},t)$. In
particular, the number density $n$, the flow velocity ${\bf u}$,  and the
pressure tensor ${\sf P}$ are obtained in terms of velocity moments of $f$,
\beq
n=\int \d {\bf v}\, f, \quad n{\bf u}=\int \d {\bf v}\, {\bf v}f,
\quad {\sf P}=m\int \d {\bf v}\, {\bf V}{\bf V}f.
\label{2.1}
\eeq
Here $m$ is the mass of a particle and ${\bf V}\equiv {\bf v}-{\bf u}$ is
the {\em peculiar\/} velocity.
The trace of ${\sf P}$ gives the hydrostatic pressure
$p=(1/d)\text{Tr}{\sf P}$, which  is related to
the temperature $T$ through the equation of state $p=nk_BT$, $k_B$ being the
Boltzmann constant.
The time evolution of $f$ is governed by the Boltzmann equation \cite{CC70}
\beq
\left(\frac{\partial}{\partial t}+v_x \frac{\partial}{\partial
x}\right)f=J[f,f],
\label{2.2}
\eeq
where $J[f,f]$ is the nonlinear Boltzmann collision operator, whose
explicit expression will be omitted here.

As happens in the uniform
shear flow \cite{DBS86,D90}, the velocity distribution function $f(x,{\bf
v},t)$ becomes spatially uniform when the velocities are referred to a
Lagrangian frame moving with the flow, i.e., $f(x,{\bf v},t)=f({\bf 
V},t)$, where ${\bf V}\equiv {\bf v}-{\bf u}(x,t)$ is the peculiar velocity.
The Boltzmann
equation (\ref{2.2}) for this flow  can then be written as
\beq
\left(\frac{\partial}{\partial \tau}-a\frac{\partial}{\partial
V_x}V_x\right)\widetilde{f}=J[\widetilde{f},\widetilde{f}],
\label{2.3}
\eeq
where
\beq
\widetilde{f}({\bf V},\tau)=\frac{n_0}{n(t)}f(x,{\bf v},t),
\quad \tau=a^{-1}\ln(1+at).
\label{2.4}
\eeq
Equation (\ref{2.3}) can be interpreted as corresponding to a
{\em homogeneous\/} gas with a velocity distribution $\widetilde{f}$ and
subject to the action of a non-conservative force $-m aV_x \widehat{\bf
x}$. Note that the density and pressure tensor associated
with $\widetilde{f}$ are $\widetilde{n}=n_0$ and
$\widetilde{P}_{ij}=(n_0/n)P_{ij}$, respectively. In fact, $\widetilde{f}$
 is proportional to the {\em probability\/} distribution of velocities.
The time variable $\tau=\int_0^t \d t' n(t')/n_0$ is a
nonlinear measure of time scaled with the number density; this variable is
unbounded even if $a<0$, since in that case
$\tau\to \infty$ when $t\to |a|^{-1}$.
It must be emphasized that Eqs.\ (\ref{2.2}) and (\ref{2.3}) are fully
equivalent in the present problem. On the other hand, Eq.\ (\ref{2.3}) has
the advantage of describing a uniform system with a constant density. The
prize to be paid is the introduction of a driving force that acts as a 
longitudinal 
``drag'' force in the case of expansion ($a>0$) and as a ``pushing'' force
in the case of condensation ($a<0$). Of course, every solution to Eq.\
(\ref{2.3}) can be mapped onto a corresponding solution to Eq.\ (\ref{2.2}).

The Boltzmann equation (\ref{2.3}) cannot be solved by analytical tools
in general. On the other hand, its associated hierarchy of moment equations
can be recursively solved in the special case of Maxwell molecules, i.e.,
particles interacting via a repulsive potential of the form $\phi(r)\propto
r^{-2(d-1)}$, in which case the collision rate is independent of the
relative velocity of the colliding particles \cite{E81}. In particular,
 Eq.\ (\ref{2.3}) yields a closed set of equations for
the elements of the pressure tensor in the case of Maxwell molecules, namely
\beq
\frac{\partial}{\partial\tau}\widetilde{p}+\frac{2a}{d}\widetilde{P}_{xx}=0,
\label{2.5.1}
\eeq
\beq
\frac{\partial}{\partial\tau}\widetilde{P}_{xx}+2a\widetilde{P}_{xx}=-\nu_0\left
(
\widetilde{P}_{xx}-\widetilde{p}\right),
\label{2.5.2}
\eeq
where
$\widetilde{p}=n_0k_BT=(1/d)\text{Tr}\widetilde{\sf P}$
and $\nu_0\propto n_0$ is a constant that plays the role of an effective 
(initial) collision
frequency.
More explicitly, $\nu_0=\widetilde{p}/\eta_{\text{NS}}$. Equation 
(\ref{2.5.1}) is not but a condition expressing the conservation 
of energy. As for Eq.\ (\ref{2.5.2}), it must be emphasized 
that it is {\em exact\/} for Maxwell molecules in our problem, i.e., no 
approximate truncation scheme (such as Grad's moment method \cite{UG99}) has 
been applied. This is a consequence of the fact that the collisional 
velocity moments of a certain degree do not involve moments of a higher 
degree in the case of the Maxwell interaction \cite{TM80}. 
The right-hand-side of Eq.\ (\ref{2.5.2}) represents the (bilinear)
collisional moment of $V_x^2$.

In this case of Maxwell
molecules the time variable $\tau$ is just proportional to the
average number of collisions per particle between $0$ and $t$, namely 
$\tau=\nu_0^{-1}\int_0^t\d t'\nu(t')$, where 
$\nu=p/\eta_{\text{NS}}=(n/n_0)\nu_0$ is the time-dependent collision 
frequency. Thus, as said in Sec.\ \ref{sec1}, every particle experiences an 
{\em infinite\/}  total number of collisions between the initial time and 
the {\em finite\/} interval $t=|a|^{-1}$ when $a<0$.
After many collision  times ($\tau\gg \nu_0^{-1}$) both 
$\widetilde{p}$ and $\widetilde{P}_{xx}$ behave as
$\widetilde{p},\widetilde{P}_{xx}\sim \exp[-{\lambda(a^*)
\nu_0\tau}]$, where $a^*\equiv a/\nu_0=A(t)/\nu(t)$ is the reduced
longitudinal rate and $\lambda(a^*)$ is the smallest root of the
quadratic equation
\beq
\lambda^2-(2a^*+1)\lambda+\frac{2}{d}a^*=0,
\label{2.6}
\eeq
i.e.,
\beq
\lambda(a^*)=a^*+\frac{1}{2}-\sqrt{\left(a^*+\frac{1}{2}\right)^2-\frac{2}{d}a^*
}
.
\label{2.7}
\eeq
The second root is obtained from (\ref{2.7}) by changing the sign of the
radical and is only relevant in the transient stage.
Consequently,
\beq
\lim_{\tau\to\infty}\frac{\widetilde{P}_{xx}}{\widetilde{p}}
=\frac{d}{2}\frac{\lambda(a^*)}{a^*}.
\label{2.8}
\eeq
In terms of the real time, one has an algebraic behavior for the
temperature, $T(t)\sim (1+at)^{-\lambda(a^*)/a^*}$.
It is important to note
that the sign of $\lambda$ is the same as  that of $a^*$. This
means that the temperature monotonically decreases in time if $a^*>0$
(expansion, drag force), while it increases if $a^*<0$ (condensation,
pushing force). In addition,  $\lambda$ is a monotonically increasing
function of $a^*$ that behaves as $\lambda \approx 2a^*+(d-1)/d$ in the
limit $a^*\to -\infty$ and as $\lambda\approx [1-(d-1)/2d a^*]/d$ in the
limit $a^*\to \infty$.

Based upon Eq.\ (\ref{1.4}),  we define  in the long-time limit a
(dimensionless) {\em nonlinear\/}
viscosity as \cite{UG99}
\beqa
\eta^*&=&\frac{d}{2(d-1)}\frac{p-P_{xx}}{A\eta_{\text{NS}}}\nonumber\\
&=&\frac{d}{2(d-1)a^*}\left(1-\frac{P_{xx}}{p}\right),
\label{2.9}
\eeqa
where we have taken into account that the bulk viscosity $\zeta_{\text{NS}}$
vanishes in a low-density gas \cite{CC70}.
Using Eq.\ (\ref{2.8}), we simply get
\beq
\eta^*(a^*)=\frac{d}{d-1}\frac{2a^*-d\lambda(a^*)}{4{a^*}^2}.
\label{2.10}
\eeq
In the three-dimensional case ($d=3$), this result coincides
with the one derived by Karlin {\em et al.} \cite{KDN97} for Maxwell
molecules by applying  the invariance principle 
under the
microscopic and macroscopic dynamics in the context of  Grad's
method.
It should also be noted that, although by a somewhat different route,
most of the above results (for $d=3$) were first derived by Galkin more than 
thirty years ago \cite{TM80,G95}.

The behavior of the nonlinear viscosity (\ref{2.10}) for small  longitudinal 
rates is
$\eta^*\approx 1-[2(d-2)/d]a^*$.
More generally, $\eta^*$ can be expressed as a series expansion in powers of 
$a^*$:
\beq
\eta^*(a^*)=1+\sum_{n=1}^\infty c_n {a^*}^n.
\label{2.10.1}
\eeq
This is just the specialization of the Chapman-Enskog expansion
\cite{CC70,DvB77}
to the simple viscous longitudinal flow. In the case of
Maxwell molecules,  $a^*=0$ is a regular point of $\eta^*$, so  the
expansion (\ref{2.10.1}) is convergent, although with a finite radius of
convergence ($|a^*|<\frac{1}{2}$) due to a branch point at
$a^*=-(d-2)/2d\pm \imath \sqrt{d-1}/d$.
The knowledge of the explicit expression of $\eta^*$,
Eq.\ (\ref{2.10}), allows one to get also its asymptotic behaviors
for large $|a^*|$; they are
$\eta^*\approx[d/2(d-1)]{a^*}^{-1}$ for $a^*>0$ and
$\eta^*\approx(d/2)|{a^*}|^{-1}$ for $a^*<0$.
This implies that $P_{xx}/p\to 0$ when $a^*\to +\infty$, while $P_{xx}/p\to 
d$
when $a^*\to -\infty$.
In the former limit all the particles tend to move along the transverse
directions, while in the latter they tend to move along the longitudinal
direction.
The shape of $\eta^*(a^*)$ for $d=3$ in the range $-2\leq a^*\leq 2$
is shown in the next Section (cf. Fig.\ \ref{fig5}).

Because of the symmetry of the problem, one expects that the heat flux is an 
irrelevant quantity \cite{UG99} that, even if it is initially different from 
zero, 
 asymptotically decays in the long-time limit. Let us analyze this point 
 more carefully. 
Taking third-degree moments in Eq.\ (\ref{2.3}), we get
\beq
\frac{\partial}{\partial 
\tau}\widetilde{q}_x+a\left(\widetilde{q}_x+2\widetilde{M}_{xxx}\right)=-
\frac{2}{3}\nu_0\widetilde{q}_x,
\label{n1}
\eeq
\beq
\frac{\partial}{\partial 
\tau}\widetilde{M}_{xxx}+3a\widetilde{M}_{xxx}=-
\nu_0\left(\frac{3}{2}\widetilde{M}_{xxx}-\frac{1}{2}\widetilde{q}_x\right),
\label{n2}
\eeq
where
\beq
\widetilde{q}_x=\frac{m}{2}\int \d {\bf V}\, V^2V_x\widetilde{f},
\quad
\widetilde{M}_{xxx}=\frac{m}{2}\int \d {\bf V}\, V_x^3\widetilde{f}.
\label{n3}
\eeq
In Eqs.\ (\ref{n1}) and (\ref{n2}) use has been made of the third-degree 
collisional moments for the three-dimensional Maxwell interaction 
\cite{TM80}.
For long collision times ($\tau\gg\nu_0^{-1}$), the third-degree moments 
behave as $\widetilde{q}_x, \widetilde{M}_{xxx}\sim 
\exp[-\omega(a^*)\nu_0\tau]$, where $\omega(a^*)\equiv 
2a^*+\frac{13}{12}-\sqrt{\left(a^*+\frac{5}{12}\right)^2-a^*}$.
If $a^*<-\frac{3}{4}+\frac{\sqrt{33}}{12}\simeq -0.271$,  $\omega$ is 
negative and then the heat flux grows in time. 
Apparently, this seems to contradict our expectation about the irrelevance 
of the heat flux in our problem.
The solution to this paradox lies in the fact that also the temperature 
grows if $a^*<0$; indeed,
what is relevant is not 
the {\em absolute\/} value of the heat flux but its value {\em 
relative\/} to the third power of the thermal velocity, namely 
$\widetilde{q}_x/mn_0(2k_BT/m)^{3/2}\sim 
\exp\{-[\omega(a^*)-\frac{3}{2}\lambda(a^*)]\nu_0\tau\}$. The difference 
$\omega(a^*)-\frac{3}{2}\lambda(a^*)$ is always positive, but goes to zero 
as $\frac{1}{12}|a^*|^{-1}$ in the limit $a^*\to -\infty$. In general, 
the smaller the value of $a^*$ the longer the transient period before 
 the heat flux, conveniently scaled with the thermal velocity, has decayed 
 to zero.

In the long-time limit, not only the {\em reduced\/} elements of the
pressure tensor $P_{ij}/p$ reach well-defined stationary values that depend
on the reduced longitudinal rate $a^*$, but the same happens with the
distribution function when properly nondimensionalized with the temperature.
This is just a statement on the validity of the ``normal'' or hydrodynamic
regime, that applies for sufficiently long times \cite{CC70,DvB77}. To be
more precise, let us introduce a reduced velocity $\bbox{\xi}$ and a reduced
distribution function $\Phi$ as
\beq
\bbox{\xi}=\left[\frac{m}{2k_BT(\tau)}\right]^{1/2}{\bf V}, \quad
\Phi(\bbox{\xi};a^*)=n_0^{-1}\lim_{\tau\to\infty}\left[\frac{2k_B
T(\tau)}{m}\right]^{d/2} \widetilde{f}({\bf V},\tau).
\label{2.11}
\eeq
In this hydrodynamic regime,
 the Boltzmann equation (\ref{2.3}) becomes (for Maxwell molecules)
\beq
\left(-a^*\frac{\partial}{\partial \xi_x}\xi_x+\frac{\lambda(a^*)}{2}
\frac{\partial}{\partial\bbox{\xi}}\cdot
\bbox{\xi}\right)\Phi=\frac{1}{\nu_0}J[\Phi,\Phi].
\label{2.12}
\eeq
As before, the first term on the left side represents a {\em driving\/}
force. The second term can be interpreted as a {\em thermostat\/}
force \cite{DBS86,EM90} that compensates for the heating ($a^*<0$) or 
cooling ($a^*>0$) effect
produced by the former.
 Equation (\ref{2.12}) yields directly the ``stationary'' values of the
(reduced) second-degree moments, namely
\beq
\langle \xi_x^2\rangle =\frac{1}{2}\frac{1}{1+2a^*-\lambda},\quad
\langle \xi_\perp^2\rangle =\frac{d-1}{2}\frac{1}{1-\lambda},
\label{2.13}
\eeq
where $\xi_\perp^2=\xi^2-\xi_x^2$. The consistency
condition $\langle \xi^2\rangle={d}/{2}$ leads again to Eq.\ (\ref{2.6}).
As noted before, $\langle \xi_x^2\rangle\to 0$ in the limit
$a^*\to +\infty$ and $\langle \xi_\perp^2\rangle\to 0$ in the opposite limit
$a^*\to -\infty$. This means that all the particles move (in the Lagrangian
frame) along directions perpendicular to the flow when $a^*\to
+\infty$, while they move parallel to the flow direction when $a^*\to
-\infty$.

\begin{figure}
\begin{center}\parbox{\textwidth}{\epsfxsize=\hsize\epsfbox{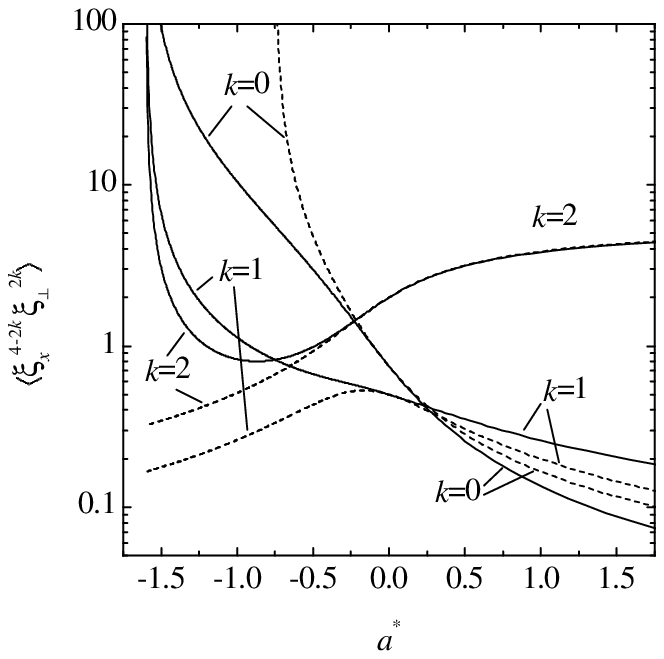}}
\caption{Plot of the reduced fourth-degree moments $\langle \xi_x^4\rangle$
($k=0$), $\langle \xi_x^2 \xi_\perp^2\rangle$ ($k=1$), and $\langle
\xi_\perp^4\rangle$ ($k=2$) as functions of the reduced longitudinal rate
$a^*$ in a three-dimensional system of Maxwell molecules. Solid
lines refer to the exact results derived from the Boltzmann equation, while
dashed lines are predictions of the BGK kinetic model.
 \label{fig1}}
\end{center}
\end{figure}

Now we can go further and consider the fourth-degree moments
$\langle \xi_x^4\rangle$, $\langle \xi_x^2\xi_\perp^2\rangle$, and $\langle
\xi_\perp^4\rangle$. Making use of the fourth-degree collisional moments for
three-dimensional Maxwell molecules \cite{TM80}, Eq.\ (\ref{2.12}) gives
rise to\beqa
2(2a^*-\lambda)\langle\xi_x^4\rangle&=&
-\case{4(2w+7)}{35}\langle\xi_x^4\rangle
+\case{3(8w-7)}{35}\langle\xi_x^2\xi_\perp^2\rangle
+\case{7-3w}{35}\langle\xi_\perp^4\rangle\nonumber\\
&&-\case{54w-91}{35}\langle\xi_x^2\rangle^2
+\case{3(36w-49)}{70}\langle\xi_x^2\rangle+\case{9(7-3w)}{70},
\label{2.14}
\eeqa
\beqa
2(a^*-\lambda)\langle\xi_x^2\xi_\perp^2\rangle&=&
-\case{144w+49}{210}\langle\xi_x^2\xi_\perp^2\rangle
+\case{8w-7}{35}\langle\xi_x^4\rangle
+\case{18w-7}{210}\langle\xi_\perp^4\rangle\nonumber\\
&&-\case{343-162w}{105}\langle\xi_x^2\rangle^2
+\case{469-216w}{140}\langle\xi_x^2\rangle
+\case{3(36w-49)}{280},
\label{2.15}
\eeqa
\beqa
-2\lambda\langle\xi_\perp^4\rangle&=&
-\case{3w+28}{35}\langle\xi_\perp^4\rangle
+\case{8(7-3w)}{105}\langle\xi_x^4\rangle
+\case{4(18w-7)}{105}\langle\xi_x^2\xi_\perp^2\rangle\nonumber\\
&&+\case{2(154w-81)}{105}\langle\xi_x^2\rangle^2
-\case{18(7-3w)}{35}\langle\xi_x^2\rangle
+\case{3(56-9w)}{70},
\label{2.16}
\eeqa
where $w\simeq 1.8731$ is the ratio of two
eigenvalues of the linearized collision operator \cite{AFP62}. The solution
of this linear set of algebraic equations is
\beqa
\langle\xi_x^4\rangle&=&\frac{3(1-3\lambda)}{4\Delta(a^*)}\left[
-1296(7-3w)\lambda^5+54(217-48w)\lambda^4-9(327w+602)\lambda^3\right.
\nonumber\\
&&\left.+33(129w+14)\lambda^2-25(75w-14)\lambda+245w\right],
\label{2.17}
\eeqa
\beqa
\langle\xi_x^2\xi_\perp^2\rangle&=&\frac{1-3\lambda}{2\Delta(a^*)}\left[
432(7-3w)\lambda^4-54(64w-91)\lambda^3\right.\nonumber\\
&&\left.+9(363w-392)\lambda^2-10(114w-35)\lambda+245w\right],
\label{2.18}
\eeqa
\beqa
\langle\xi_\perp^4\rangle&=&\frac{2}{\Delta(a^*)}\left[
324(7-3w)\lambda^4-90(6w+7)\lambda^3\right.\nonumber\\
&&\left.+9(143w+28)\lambda^2-10(114w-35)\lambda+245w\right],
\label{2.19}
\eeqa
where
\beqa
\Delta(a^*)&\equiv&(1-\lambda)^2\left[
-432(7-3w)\lambda^4-54(8w-7)\lambda^3\right.\nonumber\\
&&\left.+9(159w-56)\lambda^2-10(114w-35)\lambda+245w\right].
\label{2.20}
\eeqa
Upon writing Eqs.\ (\ref{2.17})--(\ref{2.20}) we have
made use of (\ref{2.13}) and have eliminated $a^*$ in favor of $\lambda$.
In the limit $a^*\to +\infty$, i.e., $\lambda\to \frac{1}{3}$,
we have $\langle\xi_\perp^4\rangle=9(56-9w)/2(14w+9)\simeq 5.71$ and
$\langle\xi_x^4\rangle=\langle\xi_x^2\xi_\perp^2\rangle=0$, in agreement
with a vanishing population of particles moving along the
longitudinal direction. A more interesting situation occurs in the domain of
negative $a^*$ (condensation case). As $a^*$ becomes more and more negative,
the three moments $\langle \xi_x^4\rangle$, $\langle
\xi_x^2\xi_\perp^2\rangle$, and $\langle \xi_\perp^4\rangle$ grow
monotonically and eventually diverge when $a^*$ approaches a critical value
$a_c^*\simeq -1.599$ (which corresponds to the root $\lambda_c\simeq -2.607$
of the function $\Delta$). This singular behavior of the fourth-degree
moments also occurs in the case of uniform shear flow \cite{SGBD93,SG95} and
is an indication of an algebraic high-velocity tail in the distribution
function \cite{MSG96,MSG97}. We will return to this point later on. The
moments $\langle \xi_x^4\rangle$, $\langle \xi_x^2\xi_\perp^2\rangle$, and
$\langle \xi_\perp^4\rangle$ are plotted in Fig.\ \ref{fig1}. 
While for $a^*\gtrsim 0.24$ one has $\langle \xi_x^4\rangle<\langle 
\xi_x^2\xi_\perp^2\rangle<\langle \xi_\perp^4\rangle$, the order is reversed 
for $a^*\lesssim -0.75$.

\section{Kinetic model description}
\label{sec3}
The description in the previous Section is based on the Boltzmann equation.
It has, however, two shortcomings. On the one hand, it is restricted to
Maxwell molecules. On the other hand, even for Maxwell molecules, the
explicit expression for the velocity distribution function is not known.
Both limitations are overcome if one resorts to a description less detailed
than that offered by the Boltzmann equation and employs a model kinetic
equation based on it.
The simplest and best known model kinetic equation is the one proposed
by Bhatnagar, Gross, and Krook (BGK) \cite{BGK54}.
It consists of replacing the true  collision operator $J[f,f]$ by a
single-time relaxation term of the form $-\nu (f-f_L)$, where $\nu$ is
an effective collision
frequency and
\beq
f_L=n\left(\frac{m}{2\pi
k_BT}\right)^{d/2}\exp\left(-\frac{mV^2}{2k_BT}\right)
\label{3.1}
\eeq
is the local equilibrium distribution function.
The collision frequency $\nu$ is also a
functional of $f$ through its dependence on the density and the temperature.
While the dependence on $n$ is always linear, its dependence on $T$ varies
according to the interaction potential under consideration.
For instance, in the case of repulsive potentials of the form
$\phi(r)\sim r^{-\mu}$, we simply have $\nu\propto n T^{\gamma}$ with
$\gamma=\frac{1}{2}-(d-1)/\mu$ \cite{E81}. The extreme cases correspond
to Maxwell molecules
 ($\mu=2(d-1)$, $\gamma=0$) and hard spheres ($\mu\to \infty$,
 $\gamma=\frac{1}{2}$).
For this class of repulsive potentials, the Boltzmann equation
(\ref{2.3}) is modeled as \beq \left(\frac{\partial}{\partial
\tau}-a\frac{\partial}{\partial
V_x}V_x\right)\widetilde{f}=-\nu_0(T/T_0)^\gamma(\widetilde{f}-\widetilde{f}_L),
\quad \widetilde{f}_L=(n_0/n)f_L, \label{3.2} \eeq where, as
before, the subscript  $0$ denotes  initial values.
Since the BGK model contains a single parameter ($\nu$), it is unable to 
reproduce  the correct Boltzmann values of both the shear 
viscosity an the thermal conductivity coefficients  simultaneously. In our 
problem, however, only the shear viscosity is relevant and thus the 
effective collision frequency $\nu$ can be chosen as 
$\nu=p/\eta_{\text{NS}}$, so that the exact NS viscosity is recovered.

\subsection{Maxwell molecules}
\label{subsec3.1}
In this subsection we
specialize to Maxwell molecules ($\gamma=0$). In that case, the evolution
equations for the elements of the pressure tensor are {\em identical\/} to
those already derived from the Boltzmann equation, Eqs.\ (\ref{2.5.1}) and 
(\ref{2.5.2}),
provided that the BGK collision frequency $\nu_0$ is identified with the one
arising from the Boltzmann equation.
As a consequence, the nonlinear viscosity is again given by Eqs.\
(\ref{2.10}) and (\ref{2.7}).
However, velocity moments of degree higher than two no longer coincide in both
descriptions. The main advantage of the BGK equation is that it lends itself
to an exact solution at the level of the distribution function. This
solution is expected to provide a fair description of the true distribution
at least for velocities smaller than or of the order of the thermal velocity
$(2k_BT/m)^{1/2}$.

The general solution of Eq.\ (\ref{3.2}) for Maxwell molecules is
\beqa
\widetilde{f}({\bf V},\tau)&=&
e^{-\nu_0\tau} e^{a\tau\partial_{V_x}V_x}
\widetilde{f}({\bf V},0)\nonumber\\
&&+\nu_0\int_0^{\tau}\d \tau'\,
e^{-\nu_0(\tau-\tau')}
e^{a(\tau-\tau')\partial_{V_x}V_x}\widetilde{f}_L({\bf
V},\tau'),
\label{3.3}
\eeqa
where the action of the operator $\exp(a \tau \partial_{V_x}V_x)$ is
\beq
e^{a \tau \partial_{V_x}V_x}F(V_x)=e^{a\tau}F(e^{a\tau}V_x).
\label{3.4}
\eeq
We will focus on the long-time distribution function, which
becomes independent of the choice of the initial distribution 
$\widetilde{f}({\bf V},0)={f}({\bf V},0)$. To that end,
it is convenient to work with the reduced quantities (\ref{2.11}). The BGK
counterpart of Eq.\ (\ref{2.12}) is
\beq
\left(-a^*\frac{\partial}{\partial \xi_x}\xi_x+
\frac{\lambda(a^*)}{2}\frac{\partial}{\partial\bbox{\xi}}\cdot
\bbox{\xi}\right)\Phi=-\Phi+\pi^{-d/2}e^{-\xi^2},
\label{3.5}
\eeq
whose solution is
\beq
\Phi(\bbox{\xi};a^*)=\pi^{-d/2}\int_0^\infty \d s\,
\exp\left[-\left(1-a^*+\frac{d}{2}\lambda\right)s
-e^{-\lambda s}\left(e^{2a^*s}\xi_x^2+\xi_\perp^2\right)\right].
\label{3.6}
\eeq
Of course, the same result is obtained by taking the limit
$\nu_0\tau\to \infty$ in Eq.\ (\ref{3.3}).
Equation (\ref{3.6}) also allows us to get the moments
$\langle\xi_x^{2k_1}\xi_\perp^{2k_2}\rangle$. A simple calculation yields
\beq
\langle\xi_x^{2k_1}\xi_\perp^{2k_2}\rangle=
\frac{\Gamma(k_1+\frac{1}{2})\Gamma(k_2+\frac{d-1}{2})}{\Gamma(\frac{1}{2})
\Gamma(\frac{d-1}{2})}\left[1+k_1(2a^*-\lambda)-k_2\lambda\right]^{-1}
\label{3.9}
\eeq
if $1+k_1(2a^*-\lambda)-k_2\lambda>0$, being $\infty$
otherwise. The sign of $2a^*-\lambda=(d-1)\lambda/(1-d\lambda)$ is the same
as that of $a^*$. Thus the moment
$\langle\xi_x^{2k_1}\xi_\perp^{2k_2}\rangle$ diverges when $k_1$ is
sufficiently low and $k_2$ is sufficiently high in the case $a^*>0$, while
it diverges when $k_1$ is sufficiently high and $k_2$ is sufficiently low
in the opposite case $a^*<0$. More specifically,
$\langle\xi_\perp^{2k}\rangle\to \infty$ for $a^*>0$ if $k\geq
1/\lambda(a^*)>d$; analogously,
$\langle\xi_x^{2k}\rangle\to \infty$ for $a^*<0$ if
$k\geq 1/|2a^*-\lambda|>d/(d-1)$.
In the particular case of the fourth-degree moments, $\langle \xi_x^2
\xi_\perp^2\rangle$ and $\langle \xi_\perp^4\rangle$ remain finite but
$\langle \xi_x^4\rangle$ diverges if $\lambda$ is equal to or smaller
than a critical value $\lambda_c=-1/(d-2)$, i.e., if $a^*\leq
a^*_c=-d/4(d-2)$.
This behavior is reminiscent of the one observed in Sec.\ \ref{sec2} from
the Boltzmann equation, although there are two main differences: (i) in the
case
 of the Boltzmann equation the three fourth-degree moments (i.e., not only
 $\langle \xi_x^4\rangle$) diverge and (ii) that happens for a larger
 departure from equilibrium ($a^*_c\simeq -1.599$ versus $a^*_c=-0.75$ for 
 $d=3$).
The dependence of the three moments on $a^*$, as predicted by the BGK model,
is shown in Fig.\ \ref{fig1}. It can be observed  a good agreement with
the results obtained from the Boltzmann equation in the region $a^*\gtrsim
-0.3$, especially in the case of $\langle \xi_\perp^4\rangle$. For
longitudinal rates $a^*\lesssim -0.3$, however, the deviations become
important.

It is remarkable that in this viscous longitudinal problem the BGK model is
able to capture, at least at a qualitative level, the existence of diverging
moments for values of $|a^*|$ sufficiently large. In the case of the uniform
shear flow, however, all the moments predicted by the BGK equation are
finite \cite{SB91,GS95}, in contrast to the scenario arising from the
Boltzmann equation \cite{SGBD93,SG95,MSG96,MSG97}. The origin of
diverging moments can be traced back to the existence of a high-velocity
tail in the distribution function. To clarify this point, let us consider
the two marginal distribution functions
\beqa
\varphi_\|(\xi_x;a^*)&\equiv&\int \d \bbox{\xi}_\perp \,\Phi(\bbox{\xi};a^*)
\nonumber\\
&=&\pi^{-1/2}\int_0^\infty \d s\,
\exp\left[-\left(1-a^*+\frac{1}{2}\lambda\right)s
-e^{(2a-\lambda) s}\xi_x^2\right],
\label{3.7}
\eeqa
\beqa
\varphi_\perp(\bbox{\xi}_\perp;a^*)&\equiv&\int_{-\infty}^\infty \d {\xi}_x
\,\Phi(\bbox{\xi};a^*)
\nonumber\\
&=&\pi^{-(d-1)/2}\int_0^\infty \d s\,
\exp\left[-\left(1+\frac{d-1}{2}\lambda\right)s
-e^{-\lambda s}\xi_\perp^2\right],
\label{3.8}
\eeqa
where $\bbox{\xi}_\perp\equiv \bbox{\xi}-\xi_x\widehat{\bf x}$ is the
transverse velocity vector.
A simple change of variable in
Eq.\ (\ref{3.7}) gives
\beq
\varphi_\|(\xi_x;a^*)=\pi^{-1/2}
\frac{F_\|(a^*,\xi_x^2)}{|2a^*-\lambda|}
\xi_x^{-2\beta_\|},\quad
\beta_\|(a^*)\equiv\frac{1}{2}-\frac{1}{2a^*-\lambda},
\label{3.10}
\eeq
where
\beq
F_\|(a^*,\xi_x^2)=
\left\{
\begin{array}{ll}
\Gamma\left(\beta_\|,\xi_x^2\right), & a^*>0,\\
\Gamma\left(\beta_\|\right)-\Gamma\left(\beta_\|,\xi_x^2\right), & a^*<0.
\end{array}
\right.
\label{3.13}
\eeq
Here,
\beq
\Gamma(\beta,x)=\int_x^\infty \d y\, y^{\beta-1}e^{-y}
\label{3.11}
\eeq
is the incomplete gamma function \cite{AS72}.
Analogously,
\beq
\varphi_\perp(\bbox{\xi}_\perp;a^*)=\pi^{-(d-1)/2}
\frac{F_\perp(a^*,\xi_\perp^2)}{|\lambda|}\xi_\perp^{-2\beta_\perp},\quad
\beta_\perp(a^*)\equiv\frac{d-1}{2}+\frac{1}{\lambda},
\label{3.12}
\eeq
where
\beq
F_\perp(a^*,\xi_\perp^2)=
\left\{
\begin{array}{ll}
\Gamma\left(\beta_\perp\right)-\Gamma\left(\beta_\perp,\xi_\perp^2\right), &
a^*>0,\\
\Gamma\left(\beta_\perp,\xi_\perp^2\right), & a^*<0.
\end{array}
\right.
\label{3.14}
\eeq
From Eqs.\ (\ref{3.10}) and (\ref{3.13}) it follows that
$\varphi_\|\sim \xi_x^{-2\beta_\|}$ in the limit $\xi_x^2\to\infty$ if
$a^*<0$, and so  $\langle\xi_x^{2k}\rangle\to \infty$ if $k\geq
\beta_\|-\frac{1}{2}=|2a^*-\lambda|^{-1}$.
Similarly,
$\varphi_\perp\sim \xi_\perp^{-2\beta_\perp}$ in the limit
$\xi_\perp^2\to\infty$ if $a^*>0$
and then
$\langle\xi_\perp^{2k}\rangle\to \infty$ if $k\geq
\beta_\perp-\frac{d-1}{2}=\lambda^{-1}$.
It is interesting to note that the  exponents $\beta_\|$ and
$\beta_\perp$ remain finite in the limit of infinite $|a^*|$:
$\lim_{a^*\to\infty}\beta_\perp(a^*)=
(d-1)\lim_{a^*\to-\infty}\beta_\|(a^*)=(3d-1)/2$ and
$\lim_{a^*\to-\infty}\beta_\perp(a^*)=
(d-1)\lim_{a^*\to\infty}\beta_\|(a^*)=(d-1)/2$.

Apart from an algebraic high-velocity tail, the distribution function
may exhibit a singular behavior in the opposite limit of
vanishing velocities. Equation (\ref{3.6}) shows that
$\lim_{{\bbox \xi}\to{\bf 0}}\Phi(\bbox{\xi};a^*)=\infty$ if
$1-a^*+\frac{d}{2}\lambda(a^*)\leq 0$, which corresponds to $a^*\leq
-[\sqrt{d(3d-2)}-d+2]/2(d-1)$ and $a^*\geq
[\sqrt{d(3d-2)}+d-2]/2(d-1)$.
A similar phenomenon of  overpopulation of ``rest'' particles occurs in the
uniform shear flow state \cite{SB91}.
Again, it is useful to consider the marginal distributions in the analysis
of this effect.  In the case of $\varphi_\|$ the divergence happens when
both $\beta_\|$ and $a^*$ are positive, i.e., for $a^*>3d/(3d-1)$, while in
the case of $\varphi_\perp$ the singular behavior takes place when
 $\beta_\perp>0$ and $a^*<0$,  i.e., for $a^*<-d(d+1)/(d-1)(3d-1)$.
By using the properties \cite{SB91,AS72}
\beq
\lim_{x\to 0^+}\Gamma(\beta,x)=\left\{
\begin{array}{ll}
\Gamma(\beta)-\beta^{-1}x^\beta,&\beta>0,\\
-\beta^{-1}x^\beta,&\beta<0,\\
-\ln x,&\beta=0,
\end{array}
\right. \label{3.15} \eeq one gets from Eqs.\ (\ref{3.10}) and
(\ref{3.12}) the following asymptotic behaviors: \beq
\lim_{\xi_x^2\to 0}\varphi_\|(\xi_x;a^*)=\left\{
\begin{array}{ll}
\pi^{-1/2}\left(1-a^*+\frac{1}{2}\lambda\right)^{-1}, &
a^*<\frac{3d}{3d-1},\\
\pi^{-1/2}\frac{\Gamma(\beta_\|)}{2a^*-\lambda}\xi_x^{-2\beta_\|},
&a^*>\frac{3d}{3d-1},\\
\pi^{-1/2}\ln |\xi_x|^{-1},&a^*=\frac{3d}{3d-1},
\end{array}
\right.
\label{3.16}
\eeq
\beq
\lim_{\xi_\perp^2\to 0}\varphi_\perp({\bbox \xi}_\perp;a^*)=\left\{
\begin{array}{ll}
\pi^{-(d-1)/2}\left(1+\frac{d-1}{2}\lambda\right)^{-1}, &
a^*>-\frac{d(d+1)}{(d-1)(3d-1)},\\
\pi^{-(d-1)/2}\frac{\Gamma(\beta_\perp)}{|\lambda|}\xi_\perp^{-2\beta_\perp},&a^
*<
-\frac{d(d+1)}{(d-1)(3d-1)},\\
\pi^{-(d-1)/2}(d-1)\ln |\xi_\perp|^{-1},&a^*=-\frac{d(d+1)}{(d-1)(3d-1)}.
\end{array}
\right.
\label{3.17}
\eeq

\begin{figure}
\begin{center}\parbox{\textwidth}{\epsfxsize=\hsize\epsfbox{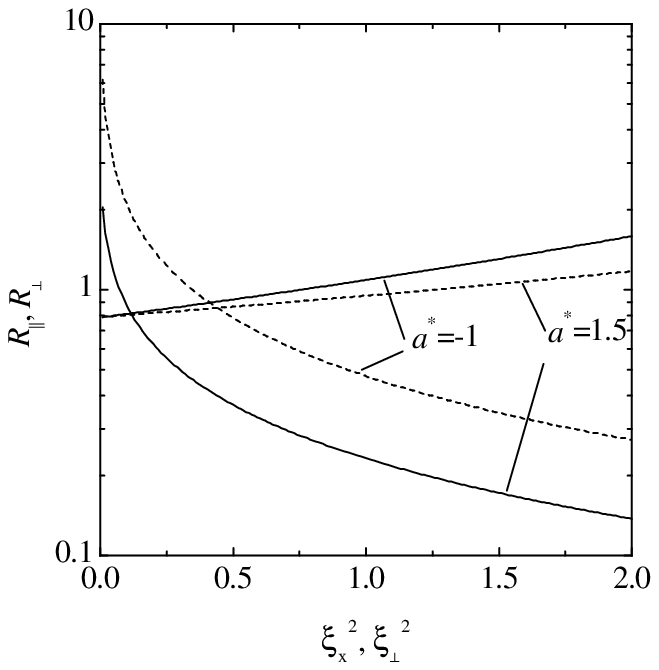}}
\caption{Marginal velocity distributions normalized with respect to local
equilibrium, $R_\|(\xi_x)$ (solid lines) and $R_\perp({\bbox \xi}_\perp)$
(dashed lines), for $a^*=-1$ and $a^*=1.5$, as predicted by the BGK kinetic
model for a three-dimensional system of Maxwell molecules.
 \label{fig2}}
\end{center}
\end{figure}

Figure \ref{fig2} shows the ratios with respect to local
equilibrium $R_\|(\xi_x;a^*)=\varphi_\|(\xi_x;a^*)/\varphi_\|(\xi_x;0)$ and
$R_\perp({\bbox \xi}_\perp;a^*)=\varphi_\perp({\bbox
\xi}_\perp;a^*)/\varphi_\perp({\bbox \xi}_\perp;0)$ for $a^*=-1$ and
$a^*=1.5$ in the three-dimensional case. It can be observed that at $a^*=-1$
($a^*=1.5$) the function $\varphi_\|$ ($\varphi_\perp$) develops a
high-velocity tail, while the function $\varphi_\perp$ ($\varphi_\|$)
diverges as the velocity vanishes.

\subsection{Repulsive potentials. Hard spheres}
\label{subsec3.2}
Now we consider more general repulsive potentials characterized by
$\gamma>0$. The BGK model for our problem is given by Eq.\ (\ref{3.2}). Its
general solution is
\beqa
\widetilde{f}({\bf
V},\tau)&=&e^{-s(\tau)}e^{a\tau
\partial_{V_x}V_x} \widetilde{f}({\bf V},0)\nonumber\\
&&+\nu_0
\int_0^{\tau}\d\tau'\,\left[\frac{T(\tau')}{T_0}
\right]^\gamma
e^{-[s(\tau) -s(\tau')]}e^{a(\tau -\tau')
\partial_{V_x}V_x} \widetilde{f}_L({\bf V},\tau'),
\label{3.18}
\eeqa
where
\beq
s(\tau)=\nu_0\int_0^{\tau}\d\tau'\,
\left[\frac{T(\tau')}{T_0}\right]^\gamma
\label{3.19}
\eeq
is the number of collisions per particle.
In the limit $\gamma\to 0$, $s(\tau)=\nu_0\tau$ and Eq.\
(\ref{3.18}) reduces to Eq.\ (\ref{3.3}). On the other hand, for $\gamma>0$
Eq.\ (\ref{3.18}) is not closed since it requires the knowledge of the time
dependence of the temperature. By multiplying both sides of Eq.\
(\ref{3.18}) by $V^2$ and integrating over velocity one can get a closed
integral equation for $T(\tau)$. Such an equation is however quite
involved and then it is more transparent to work directly with the evolution
equation itself, Eq.\ (\ref{3.2}). From this equation it is straightforward
to find that the
evolution of $\widetilde{p}$ and $\widetilde{P}_{xx}$ is still given by Eqs.\
(\ref{2.5.1}) and (\ref{2.5.2}), except that now $\nu_0$ is replaced by 
a time-dependent
collision frequency
$\nu_0[\widetilde{p}(\tau)/p_0]^\gamma$.
Therefore, the pressure $\widetilde{p}=n_0 k_B T$ obeys a {\em nonlinear\/}
second-order differential equation
\beq
\frac{\partial^2}{\partial \tau^2}\widetilde{p}+
\left[2a+\nu_0
\left(\frac{\widetilde{p}}{p_0}\right)^\gamma\right]\frac{\partial}{\partial
\tau}\widetilde{p}+\frac{2}{d}a\nu_0
\left(\frac{\widetilde{p}}{p_0}\right)^\gamma\widetilde{p},
\label{3.20}
\eeq
subject to the initial conditions $\widetilde{p}(0)=p_0$, $\left.\partial
\widetilde{p}/\partial
\tau\right|_{\tau=0}=-(2a/d)P_{xx}(0)$.
For asymptotically long times ($\tau\to \infty$), the solution of
Eq.\ (\ref{3.20}) behaves as $\widetilde{p}(\tau)\sim
\exp(-2a\tau/d)$ if $a<0$ and as $\widetilde{p}(\tau)\sim
(1+\gamma\nu_0\tau/d)^{-1/\gamma}$ if $a>0$.
In both cases the accumulated number of collisions $s(\tau)$, Eq.\ 
(\ref{3.19}), goes to infinity as $\tau\to\infty$. In particular, if 
$a<0$, the typical number of collisions per particle during the finite 
interval $0\leq t\leq |a|^{-1}$ becomes infinite, a property already seen in 
Sec.\ \ref{sec2} for Maxwell molecules.

In order to get the
nonlinear viscosity as a function of the longitudinal deformation rate, we
must work with the reduced rate
$a^*=A/\nu=a/\nu_0(\widetilde{p}/p_0)^\gamma$ rather than with the time
variables $\tau$ or $t$. Note that
$\lim_{\tau\to\infty}a^*(\tau)=\infty$ if $a>0$,
while $\lim_{\tau\to\infty}a^*(\tau)=0$ if $a<0$.
With this change of variable one has
\beq
\frac{\partial}{\partial a^*}\widetilde{p}=-\frac{\widetilde{p}}{\gamma
a^*},
\label{3.21a}
\eeq
\beq
\frac{\partial}{\partial
a^*}\widetilde{P}_{xx}=-\frac{d\widetilde{p}}{\gamma a^*}
\left[1+\frac{1}{2a^*}\left(1-\frac{\widetilde{p}}{\widetilde{P}_{xx}}\right)
\right].
\label{3.21b}
\eeq
This gives rise to the following ordinary differential equation for the
reduced nonlinear viscosity defined in Eq.\ (\ref{2.9}):
\beq
2\gamma{a^*}^2\left(1-2\frac{d-1}{d}a^*\eta^*\right)
\frac{\partial\eta^*}{\partial a^*}
+4\frac{d-1}{d}(1-\gamma){a^*}^2{\eta^*}^2+
\left[d+2(d-2+\gamma)a^*\right]\eta^*-d=0.
\label{3.22}
\eeq
For small $a^*$ the solution is $\eta^*\approx 1-[2(d-2+\gamma)/d]a^*$,
while the asymptotic behavior of $\eta^*$ for large
$|a^*|$ is  $\eta^*\approx [d/2(d-1)]{a^*}^{-1}(1-{a^*}^{-1}/2)$
for $a^*>0$ and $\eta^*\approx (d/2)|a^*|^{-1}[1-|a^*|^{-1}/2(1+\gamma)]$
for $a^*<0$. The leading terms are independent of $\gamma$ and correspond
to $\lim_{a^*\to +\infty}P_{xx}/p=0$ and $\lim_{a^*\to -\infty}P_{xx}/p=d$,
respectively.
Interestingly enough, Eq.\ (\ref{3.22}), particularized to $d=3$, is
equivalent to the one derived by Karlin {\em et al.} \cite{KDN97} from
 their invariance principle and Grad's method. In other words, the
invariance principle under the microscopic and macroscopic dynamics is an
approximation that, at least in this problem, yields the same  nonlinear
viscosity as the one predicted by the BGK model. The latter approach,
nevertheless, has the advantages of being conceptually simpler and providing
the full velocity distribution function.

\begin{figure}
\begin{center}\parbox{\textwidth}{\epsfxsize=\hsize\epsfbox{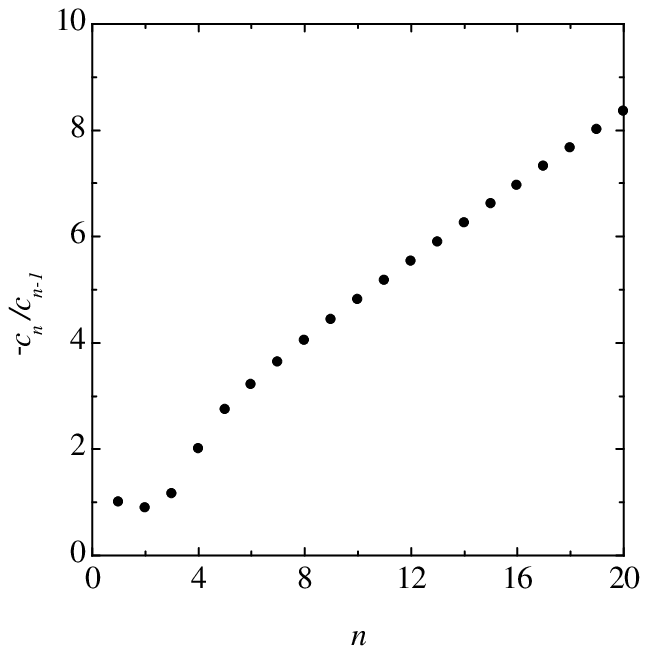}}
\caption{Ratio -$c_{n}/c_{n-1}$ between successive coefficients in
the Chapman-Enskog expansion of the nonlinear viscosity for a
three-dimensional system of hard spheres, according to the BGK
kinetic model.
 \label{fig3}}
\end{center}
\end{figure}

It must  be noted that Eq.\ (\ref{3.22}) possesses as many solutions
as particular initial conditions. Each particular solution is
specified by assigning a given value of the viscosity $\eta^*$ at
the initial (reduced) longitudinal rate $a_0^*=a/\nu_0$. This
situation is analogous to the one discussed in Ref.\ \cite{SB91}
for the uniform shear flow case. Since the irreversible time
evolution of the system leads to a monotonic increase of $a^*$,
Eq.\ (\ref{3.22}) must be solved for $a^*\geq a_0^*$. For positive
rates, this implies the range $0< a_0^*\leq a^*$, but for negative
rates the range is $a_0^*\leq a^*<0$. All the particular
solutions, however, tend towards a special solution (the
hydrodynamic one) for sufficiently long times, i.e., for $a^*\gg
a_0^*$ if $a_0^*>0$ and for $|a^*|\ll |a_0^*|$ if $a_0^*<0$.
 This special solution representing the
hydrodynamic or normal regime can be identified in principle by the
Chapman-Enskog series (\ref{2.10.1}). Insertion into the differential
equation (\ref{3.22}) yields the following recurrence formula,
\beq
c_n=-2\frac{d-2+n\gamma}{d}c_{n-1}-4\frac{d-1}{d^2}\sum_{m=0}^{n-2}c_m
c_{n-2-m}\left[1-(n-1-m)\gamma\right].
\label{3.23}
 \eeq
This equation shows that the coefficient $c_n$ is a polynomial in $\gamma$
of degree $n$.
The first four coefficients in the case of a
three-dimensional system of hard spheres ($d=3$, $\gamma=\frac{1}{2}$)
are $c_1=-1$, $c_2=\frac{8}{9}$, $c_3=-\frac{28}{27}$, and
$c_4=\frac{56}{27}$.
Further computation of the coefficients shows that, except in the case of
Maxwell molecules ($\gamma=0$), the expansion (\ref{2.10.1}) is {\em only
asymptotic}. For large $n$, the magnitude of the coefficients $c_n$ grow so
rapidly that the second term on the right-hand-side of Eq.\ (\ref{3.23}) can
be neglected, so  $c_n/c_{n-1}\approx -2n\gamma/d$. The ratio
$-c_n/c_{n-1}$, $n=1$--$20$, is plotted in Fig. \ref{fig3}   for
three-dimensional
 hard spheres. The linear growth of the ratio is already apparent for
 $n\geq4$.
The divergence of the Chapman-Enskog expansion for $\gamma>0$ also takes
place in the uniform shear flow problem\cite{SB91,SBD86}.

\begin{figure}
\begin{center}\parbox{\textwidth}{\epsfxsize=\hsize\epsfbox{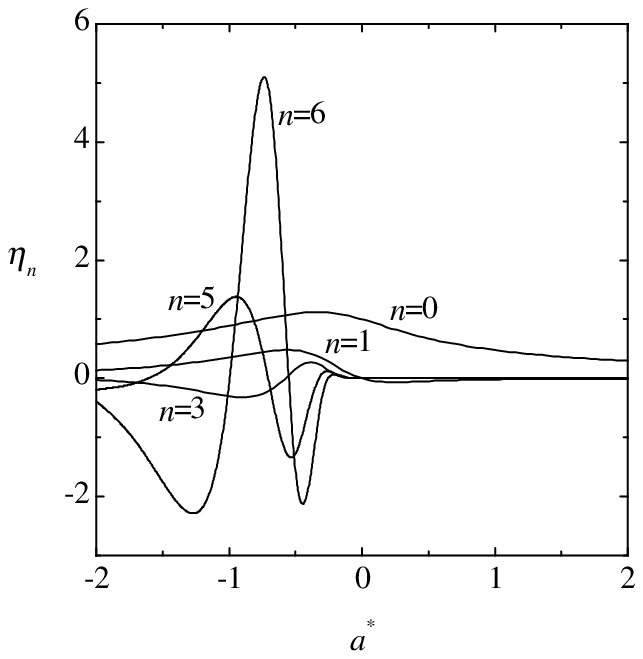}}
\caption{Longitudinal-rate dependence of the coefficients $\eta_n$,
$n=0,1,3,5,6$ of the expansion of the nonlinear viscosity in powers of the
interaction parameter $\gamma$, according to the BGK kinetic model for
three-dimensional systems.
 \label{fig4}}
\end{center}
\end{figure}

\begin{figure}
\begin{center}\parbox{\textwidth}{\epsfxsize=\hsize\epsfbox{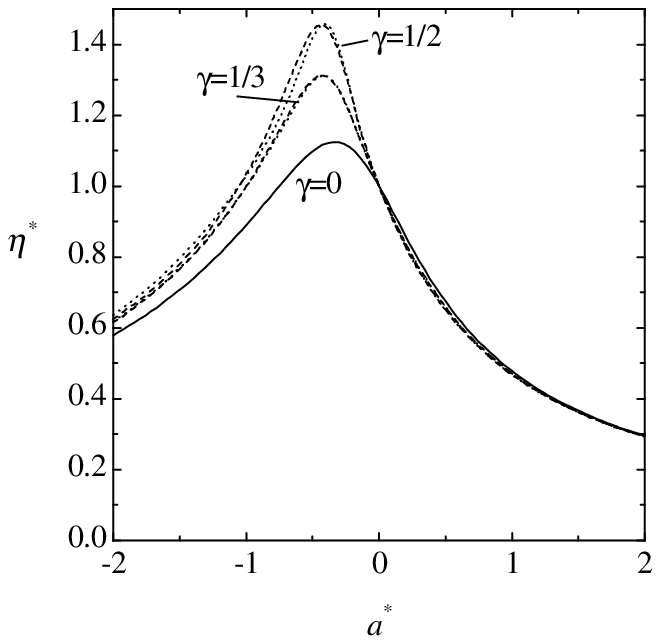}}
\caption{Nonlinear viscosity for three-dimensional systems of Maxwell
molecules ($\gamma=0$), particles interacting via an $r^{-12}$-potential
($\gamma=\frac{1}{3}$), and hard spheres ($\gamma=\frac{1}{2}$). The solid
line is the exact result derived from the Boltzmann equation and the BGK
model, while the dashed and dotted lines are the approximations $\eta^{(3)}$
and $\eta^{(4)}$, respectively, as obtained from the BGK kinetic model.
 \label{fig5}}
\end{center}
\end{figure}

Since the series (\ref{2.10.1})  is only useful if truncated and applied to
small $a^*$, a different strategy is needed to get the hydrodynamic $\eta^*$
for finite $a^*$. One possibility is to expand $\eta^*$ around the point at
infinity
(namely in powers of ${a^*}^{-1}$). Such an expansion proved to be
convergent in the case of uniform shear flow  for shear
rates larger than a certain value \cite{SB91,SBD86}. From a practical
point of view, however, this method is not very convenient because many
terms would need to be retained in order to get reliable results in the
range of interest (say $|a^*|\sim 1$), even if the expansion converges.
A second possibility is to solve numerically  the differential equation
(\ref{3.22}) with the boundary conditions $\lim_{a^*_0\to
0^+}\eta^*(a^*_0)=1$ (for $a^*>0$) and $\lim_{a^*_0\to-\infty
}\eta^*(a^*_0)=0$ (for $a^*<0$). On the other hand,
 it seems more convenient to follow the approach proposed in Ref.\
\cite{KDN97}, which consists of representing the nonlinear viscosity as
an expansion in powers of the interaction parameter $\gamma$:
 \beq
\eta^*(a^*)=\lim_{N\to\infty}{\eta}^{(N)}(a^*),\quad {\eta}^{(N)}(a^*)=
\sum_{n=0}^N \eta_n({a^*})\gamma^n,
\label{3.24}
\eeq
where $\eta_0(a^*)$ is the nonlinear viscosity for Maxwell molecules, Eq.\
(\ref{2.10}).
Since the Chapman-Enskog coefficients $c_n$ are polynomials in $\gamma$, the
series (\ref{3.24}) can be interpreted as a rearrangement of the series
(\ref{2.10.1}). In other words, if we write
\beq
c_n=\sum_{m=0}^n c_{nm}\gamma^m,
\label{3.24.1}
\eeq
then
\beq
\eta_{n}(a^*)=\sum_{m=n}^{\infty} c_{mn} {a^*}^m.
\label{3.24.2}
\eeq
As a consequence, the truncated series $\eta^{(N)}(a^*)$ is exact through
order ${a^*}^N$, i.e., $\eta^*(a^*)-\eta^{(N)}(a^*)={\cal O}({a^*}^{N+1})$.
Insertion of Eq.\ (\ref{3.24}) into Eq.\ (\ref{3.22}) yields the following
recurrence formula
\beqa
\eta_n({a^*})&=&-\frac{2a}{d+2(d-2)a^*+8\frac{d-1}{d}{a^*}^2\eta_0(a^*)}
\left\{\left[1-2\frac{d-1}{d}a^*\eta_0(a^*)\right]\left[\eta_{n-1}(a^*)+
a^*\eta_{n-1}'(a^*)\right]\right.\nonumber\\
&&\left.-2\frac{d-1}{d}a^*\sum_{m=1}^{n-1}\eta_m(a^*)\left[\eta_{n-1-m}(a^*)
+a^*
\eta_{n-1-m}'(a^*)-\eta_{n-m}(a^*)\right]\right\},
\label{3.25}
\eeqa
where the prime denotes a derivative with respect to $a^*$.
The coefficient $\eta_1$ for $d=3$ was the only one considered
 by Karlin {\em et al.} \cite{KDN97,KDN98,note}. The coefficients $\eta_n$
 for
$n=0,1,3,5,6$ and $d=3$ are plotted in Fig.\ \ref{fig4}. Up to $n=3$ the
coefficients remain small, but the magnitude of $\eta_5$ and, especially,
that of $\eta_6$ reach rather high values, thus suggesting the asymptotic
character of the expansion (\ref{3.24}), at least for negative $a^*$. This
is not surprising if one takes into account that, while all the truncated
series  $\eta^{(N)}(a^*)$ are regular at $a^*=0$, the full viscosity
$\eta^*(a^*)$ is singular at $a^*=0$.
Notwithstanding this, since the maximum value of $\gamma$ is
$\gamma=\frac{1}{2}$, it turns out that the functions $\eta^{(N)}(a^*)$ with
$N=3$ or $N=4$ can be considered as rather good approximations of
$\eta^*(a^*)$. This is quite apparent in Fig.\ \ref{fig5}, where
$\eta^{(3)}(a^*)$ and $\eta^{(4)}(a^*)$ practically overlap in the case
$\gamma=\frac{1}{3}$ (corresponding to a repulsive potential with
$\mu=12$) and are hardly distinguishable in the case of
$\gamma=\frac{1}{2}$ (hard spheres).  Figure
\ref{fig5} also shows that the nonlinear viscosity is almost insensitive to
the interaction model in the case of an expansion ($a^*>0$). On the other
hand, when the physical situation corresponds to a condensation of the gas
($a^*<0$), the hardness of the repulsion plays a relevant role, especially
around the maximum ($a^*\approx -0.4$) \cite{note2}.

Before closing this Section, it is worthwhile noting that Uribe and 
Garc\'{\i}a-Col\'{\i}n \cite{UG99} used  Grad's  (nonlinear) moment method 
to get an expression for $\eta^*(a^*)$ that is dramatically at odds with the 
results obtained in this paper for $a^*<0$. According to  their 
results, $\eta^*(a^*)$ monotonically increases as $a^*$ becomes more 
and more negative and finally reaches a plateau $\eta^*\to 49$ in the limit 
$a^*\to -\infty$. However, these results are strongly inconsistent with the 
physical condition $p=[P_{xx}+(d-1)P_{yy}]/d\geq P_{xx}/d$, which implies 
[cf.\ Eq.\ (\ref{2.9})] that $\eta^*(a^*)\leq (d/2) |a^*|^{-1}$ if $a^*<0$. 
Since $\eta^*$ has  an upper bound that goes to zero in the limit 
$a^*\to -\infty$, then $\lim_{a^*\to -\infty} \eta^*(a^*)=0$ necessarily. On 
the other hand, from Eqs.\ (45) or (47) of Ref.\ \cite{UG99} it follows that 
$p<P_{xx}/3$ (i.e, $P_{yy}<0$) if $a^*<-\frac{5}{14}$. These inconsistencies 
of the results derived in Ref.\ \cite{UG99} are likely associated with the 
assumption of a stationary situation in this compressible flow.

\section{Conclusions}
\label{sec4}
This paper has dealt with a simple viscous longitudinal flow characterized
by an unsteady velocity profile $u_x(x,t)=a x/(1+at)$ and a uniform density
$n(t)=n_0/(1+at)$. The situation with $a>0$ corresponds to an expansion of
the gas, while the case $a<0$ describes a condensation phenomenon. By using
kinetic theory tools (Boltzmann equation and BGK kinetic model), exact
results have been derived for the generalized or nonlinear viscosity, the
velocity moments, and the velocity distribution function. The following
points summarize the main conclusions of the present study.
\begin{itemize}
\item
By an adequate change of velocity and time variables, the problem is seen to
be formally equivalent to that of a uniform gas with a steady density, in
which the particles are under the action of a longitudinal driving force
${\bf F}=-m a V_x \widehat{\bf x}$. According to this viewpoint, the
particles are decelerated or accelerated along the longitudinal direction,
depending on the sign of $a$. As a consequence, the temperature
monotonically decreases in time if $a>0$, while it increases if $a<0$.
\item
The relative difference between the normal stress $P_{xx}$ and the
hydrostatic pressure $p$ for long collision times is characterized by a
(dimensionless) viscosity coefficient $\eta^*(a^*)$, which is a nonlinear
function of the longitudinal deformation rate relative to an effective
collision frequency. The (Chapman-Enskog) expansion of $\eta^*$ in powers of
$a^*$ is, in general, only asymptotic. An exception is provided by Maxwell
molecules, in which case the expansion converges for $|a^*|<\frac{1}{2}$.
\item
A thinning effect is present for $a^*>0$, i.e., $\eta^*$ monotonically
decreases as $a^*$ increases. For $a^*<0$, however, $\eta^*$ starts
increasing with $|a^*|$ (thickening effect), reaches a maximum
(at $a^*=-\frac{1}{3}$ for Maxwell molecules and around $a^*\simeq -0.4$
for hard spheres), and then decreases for more negative longitudinal rates.
\item
In the case of an expansion ($a^*>0$), the nonlinear viscosity $\eta^*$ is
practically ``universal''. On the other hand, its behavior in the case of
condensation ($a^*<0$) is rather sensitive to the interaction potential. In
particular, the harder the potential the higher the maximum value of
$\eta^*$ (for instance, $\eta^*_{\text{max}}=1.125$ for Maxwell molecules and
$\eta^*_{\text{max}}\simeq 1.46$ for hard spheres).
\item
The results for $\eta^*$ derived from the Boltzmann equation for Maxwell
molecules and from the BGK model for more general potentials coincide with
those derived by Karlin {\em et al.} \cite{KDN97} from  Grad's method and
the application of their invariance principle under microscopic and
macroscopic dynamics. It would be interesting to explore whether such an
equivalence extends to the similar but more complicated problem of uniform
shear flow as well.
\item
The shape of the velocity distribution function (for Maxwell
molecules) has also been analyzed by scaling the velocities with
the (unsteady) thermal velocity. This gives rise to a new term in
the kinetic equation that represents a non-conservative thermostat
force that cancels the cooling ($a^*>0$) or heating ($a^*<0$)
produced by the driving force. This exact equivalence between the
free system and the thermostatted one is analogous to that taking
place in the uniform shear flow and is restricted to Maxwell
molecules (collision frequency independent of the velocity). For
other interactions the equivalence is only approximate, but yet
the thermostatted problem is worth studying by itself. It would be
quite interesting to carry out nonequilibrium  molecular dynamics
simulations of hard spheres  subject to the simultaneous action
of the driving and thermostat forces, in order to measure the
nonlinear viscosity and related phenomena in dense gases. This
would complement the extensive simulation
studies of the uniform shear flow.
\item
The exact fourth-degree (scaled) velocity moments derived from the Boltzmann
equation diverge for sufficiently negative values of the longitudinal rate
($a^*\lesssim  -1.6$).
This indicates the existence of an algebraic high-velocity tail, especially
for the longitudinal component of the velocity, for negative $a^*$.
\item
The above singular behavior of the moments is described, at least
qualitatively, by the exact solution of the BGK model.
This is a very remarkable feature, since a similar behavior showing up in
the uniform shear flow was not captured by the kinetic model.
Also, note that Grad's method  is unable to predict an
algebraic high-velocity tail, as it approximates the distribution
function by a Gaussian times a polynomial.
The BGK solution obtained in this paper predicts an algebraic tail in the
marginal distribution of longitudinal velocities if $a^*<0$ and a weaker
tail in the marginal distribution of transverse velocities if $a^*>0$.
The latter tail implies that, while all the moments of degrees equal to or
smaller than six (for a three-dimensional system) are finite when $a^*>0$,
the moments of eighth degree diverge if $a^*\geq 1.125$.
The
investigation of whether or not
this prediction is confirmed by  the Boltzmann equation will be the
subject of a separate paper.
\item
The explicit expression of the BGK velocity distribution function allows one
to unveil a different type of singular behavior that does not have, however,
a direct influence on the velocity moments since it is associated with
the limit of small velocities. More specifically, the  distribution
of vanishing  velocities diverges if the longitudinal rate is
sufficiently positive ($a^*\gtrsim 1.40$) or sufficiently negative
($a^*\lesssim -0.90$). This effect is also present in the marginal
distribution of longitudinal velocities (for $a^*\geq 1.125$) and in that of
transverse velocities (for $a^*\leq -0.75$).

\end{itemize}
\acknowledgments
The author acknowledges partial support from the DGES (Spain)
through grant No.\ PB97-1501 and from the Junta de Extremadura (Fondo Social
Europeo)
through grant No.\ IPR99C031.



\begin{references}
\bibitem{note}
Note a misprint in Eq.\ (16) of Ref.\ \protect\cite{KDN97}: The fraction 
$(3/2)$ in the numerator should be replaced by $(1-gR_0)$.
\bibitem{note2}
It is worthwhile mentioning that the approximation $\eta^{(1)}$ 
underestimates the maximum by about $9\%$ in the case of hard spheres.
\end{references}

\newpage
\newpage
\newpage
\newpage
\newpage

\end{document}